\documentclass[aps,pre,twocolumn,superscriptaddress,floatfix]{revtex4}
\usepackage{graphicx}
\usepackage[unicode,colorlinks,linkcolor=blue,citecolor=blue,urlcolor=blue]{hyperref}
\begin{document}
\title{Heat conduction in a chain of colliding particles with stiff repulsive potential}

\author{Oleg V. Gendelman}
\email[]{ovgend@tx.technion.ac.il}
\affiliation{Faculty of Mechanical Engineering, Technion -- Israel Institute of Technology,
           Haifa 32000, Israel}

\author{Alexander V. Savin}
\email[]{asavin@center.chph.ras.ru}
\affiliation{Semenov Institute of Chemical Physics, Russian Academy
of Sciences, Moscow 119991, Russia}

\date{\today}

\begin{abstract}
One-dimensional billiard, i.e. a chain of colliding particles with equal masses, is well-known
example of completely integrable system. Billiards with different particles are generically not integrable,
but still exhibit divergence of a heat conduction coefficient (HCC) in thermodynamic limit. Traditional billiard models
imply instantaneous (zero-time) collisions between the particles. We lift this condition and consider the heat
transport in a chain of stiff colliding particles with power-law potential of the nearest-neighbor
interaction. The instantaneous collisions correspond to the limit of infinite power in the
interaction potential; for finite powers, the interactions take nonzero time. This modification
of the model leads to profound physical consequence -- probability of multiple, in particular,
triple particle collisions becomes nonzero. Contrary to the integrable billiard of equal particles,
the modified model exhibits saturation of the heat conduction coefficient for large system size.
Moreover, identification of scattering events with the triple particle collisions leads to simple
definition of characteristic mean free path and kinetic description of the heat transport.
This approach allows prediction both of temperature and density dependencies for the HCC limit values.
The latter dependence is quite counterintuitive - the HCC is inversely proportional to the particle density in the chain.
Both predictions are confirmed by direct numeric simulations.
\end{abstract}
\maketitle

Microscopic description of heat conduction in dielectrics remains open and elusive problem despite
rather long history \cite{FPU, PE, jack89, GM92} and intensive research efforts over two last
decades \cite{LLP97, R1, R2, R4, R5, R6, LLP03, LLP16}.
One of most intriguing questions is a convergence of the heat conduction coefficient (HCC)
in thermodynamic limit \cite{LLP97, LLP03, LLP16}.
Common understanding achieved as a result of these efforts suggests,
that in the lattices with low-order polynomial nonlinearity (for instance,
famous Fermi-Pasta-Ulam lattice, \cite{LLP97}) the behavior of HCC strongly depends on dimensionality.
Namely, in one-dimensional lattices it diverges in thermodynamic limit as,
$L^\delta$, $0.3\le \delta\le 0.4$,
where $L$ is the size (or number of particles) in the system. For 2D, the HCC is believed
to behave as $\ln(L)$, \cite{LLP03, 2D}, and for 3D, finally, converges \cite{SATO}.
This common understanding is supported by solid theoretical arguments based on a combination
of different approaches \cite{T1, T2, T3, T4, T5}. These approaches
provide somewhat different estimations for the divergence exponent $\alpha$
(within the range of measured values for different model potentials),
but in general this part of the picture seems self-consistent.

In the same time, it has been claimed for long that in some 1D chains, for instance,
in the chain of rotators, \cite{GS00, Giardina00}
the HCC converges in the thermodynamic limit despite the momentum conservation.
More recent results of this sort, namely, the HCC convergence in Lennard-Jones (LJ) chain,
were reported in \cite{LJ-in}.  In this paper, similar convergent behavior has been claimed also
for $\alpha$-$\beta$ FPU chain and attributed to the asymmetry of the interaction potential.
This latter claim for the $\alpha$-$\beta$ FPU has been disproved
in \cite{FPU-anti}; the LJ chain has not been addressed there.

From physical point of view, the low-order polynomial nonlinearity of the FPU and similar
models arises as Taylor truncation of the complete interaction potential.
Possibility of such truncation, "self-evident"  at least
for low temperatures, seems however problematic in the thermodynamic limit; to remind, the latter
corresponds to infinite size of the system and infinite time. Any realistic physical potential
of interaction should tend to zero as the interacting atoms are at large distance --
in other terms, it should allow dissociation, like in the LJ chain. The polynomial truncation
definitely fails to describe this feature and yields instead an unphysical infinite attraction
force. In realistic system the dissociation or formation of abnormally long links between
the particles has exponentially small, but nonzero probability at low temperatures.
The polynomial truncation precludes such behavior completely. Such long links can presumably scatter
phonons quite efficiently, and thus could modify the HCC convergence properties.
Further results on the HCC convergence in many 1D models with possibility of dissociation were
reported in \cite{SK14, GS14}. The HCC convergence in systems of LJ particles and particles
with elastic shell has been observed in a number of additional studies \cite{ZSG15}.

In the same time, recent treatise on non-equilibrium hydrodynamics of the anharmonic chains
\cite{Spohn14} points on important difference between the aforementioned model of rotators and
the models similar to the FPU or LJ chains. The difference is a number of conservation laws; for the chain
of rotators, only total momentum and energy are conserved. In FPU, LJ and similar chains,
in addition, a total length of the system is conserved. This additional conservation law obviously
does not depend on the possibility of dissociation. This qualitative difference is believed
to lead to difference in the HCC convergence properties \cite{Spohn14}. From this point of view,
in the thermodynamical limit all non-integrable chains with three conservation laws mentioned
above should behave qualitatively in a similar manner and thus have the divergent HCC. From this
viewpoint, the observed convergence in the LJ chain, chain of elastic rods and similar
dissociating chains may be interpreted as finite-size effect. Such "finite-size"\ saturation
of the HCC with resumed growth for larger system sizes has been demonstrated in
$\alpha$-$\beta$ FPU \cite{FPU-anti} and in a chain of rigid particles with alternating masses
\cite{Casati}. To the best of the authors' knowledge, no "resumed"\ growth of the HCC in LJ
or similar models has been reported so far. In the same time, one should admit that any numeric
simulation in general cannot prove (or disprove) the HCC convergence in the thermodynamical
limit for any model. To be on the safe side, we would refer to the observed phenomenon as
saturation of the HCC for certain large scale of the system, without explicit claim of the
convergence. In the LJ and similar models with dissociation this saturation occurs at the scale of $10^4 - 10^5$
particles. For typical interatomic distances, such specimen will have a length of order
1-10~$\mu$m.

The HCC saturation in the LJ chain and the chain of elastic rods have one more important common
feature. The HCC behavior in the saturation regime can be interpreted in terms of simple kinetic
theory \cite{GS14}. For the chain of elastic rods, one can predict the dependence of the HCC
on the temperature and other system parameters. Similar estimations (to lesser extent) are available also
for the LJ chain. This simple kinetics seems related to observed exponential decay of the
autocorrelation of the heat flux in the saturation regime.

The claim on the HCC saturation in the dissociating chains has a profound
counterexample, or even  a group of counterexamples \cite{LLP16}. The 1D billiard of perfectly rigid colliding particles with
equal masses has obviously divergent heat conductivity. Moreover, this model is completely
integrable and therefore unable to form even linear temperature profile, when attached to thermostats.
For the point 1D billiard, this integrability is preserved even in presence of an on-site
potential \cite{GS04}. Other billiard models are not integrable,
but also exhibit divergent heat conductivity \cite{Dharrev}.

Current paper addresses this group of counterexamples. Traditional billiard models have one
important common feature -- the collisions between the particles are instantaneous,
they take zero time. Such behavior requires infinitely large interaction force.
So, potential of interaction between such particles includes vertical potential wall.
Such instantaneous collisions are apparently unphysical, since a repulsive core of any realistic
interatomic potential grows rapidly, but with finite rate at nonzero distances.
So, the realistic interparticle collision will take some finite, maybe very small, but nonzero time.
We claim that this peculiarity leads to drastic change in the transport properties of the 1D chain,
since it makes a probability of triple collisions nonzero. In the case of equal masses the double collisions,
even with finite interaction time, do not violate integrability.
The reason  is that, as a result of momentum and energy conservation, the colliding particles
with equal mass just exchange their momenta, similarly to the instantaneous collisions.
The triple collisions, however, do violate the integrability. We are going to demonstrate
that they also bring about the HCC saturation in the 1D case. So, similarly to the case
of the FPU-type chains, correction of the unphysical features of interaction potential
may lead to significant modification of the heat transport properties,
at least at the saturation mesoscale.
\begin{figure}[tb]
\includegraphics [angle=0, width=1\linewidth]{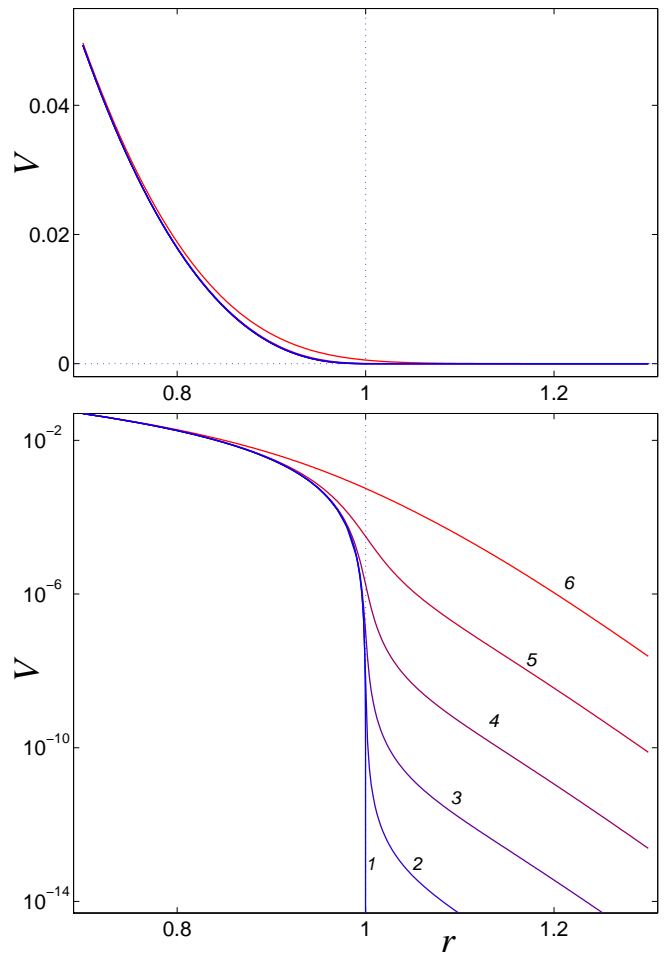}
\abovecaptionskip=20pt
\caption{(Color online)
Comparison between potential of interaction, given by formula (\ref{f1}) (curve 1)
and smoothed potential $V_h(r)$, given by (\ref{f2}), for
$h=0.000001$, 0.00001, 0.0001, 0.001, 0.01 (curves 2, 3, 4 ,5, 6), parameter $\alpha= 5/2$.
}
\label{fig01}
\end{figure}

To demonstrate that, we consider a chain of one-dimensional rods with the following purely
repulsive potential of the nearest-neighbor interaction:
\begin{eqnarray}
V(r) &=& 0,~~\mbox{for}~~r>D, \nonumber\\
V(r) &=& V_l(r)=K|D-r|^\alpha,~~\mbox{for}~~ r\le D, \label{f1}
\end{eqnarray}
Here $\alpha\ge 2$ -- parameter that governs the growth of the repulsive force, $r$
is the distance between the centers of neighbor rods; $D$ is the size of the rod.
The case $\alpha=2$ corresponds to semi-elastic rods considered in \cite{GS14}; $\alpha=5/2$
corresponds to the case of Hertzian contact. Without restricting the generality, we further use
non-dimensional parameters $D=1$, $K=1$. Then, the limit $\alpha\rightarrow\infty$
corresponds to the case of perfect instantaneous elastic collision as $r=1$.

In order to avoid numeric problems related to non-analyticity of potential (\ref{f1}) at $r=1$,
we substitute it in  the simulations by smoothed potential function
\begin{equation}
V_h(r)=2^{-\alpha}\left[\sqrt{\rho^2+hf(\rho)}-\rho\right]^{\alpha},~~\rho=r-1,
\label{f2}
\end{equation}
where function $f(\rho)=1/(1+5\rho^2)^6$, parameter $h>0$. In  the limit $h\rightarrow 0$ the smoothed
potential (\ref{f2}) tends to non-analytic potential (\ref{f1}).
Comparison between exact and smoothed potentials of interaction is presented in Fig.~\ref{fig01}
for different values of the smoothing parameter $h$.

We perform traditional numeric simulation of the equilibrium heat transport
in one-dimensional model and consider a segment of length $L$ parallel to $x$ axis. We pack $N=p(L-1)+1$ rods
along this segment, where $p$ ($0<p<1$) stands for the packing "density"\ of the chain.
Fixed boundary conditions are imposed on both ends of the chain, i.e. $x_1\equiv 0$, $x_N\equiv (N-1)a$,
where $a=1/p$ stands for the period of the unperturbed chain. Fixed boundaries enforce
the density conservation. The disks $1<n<N$ are then restricted to move in
$x$ direction. Hamiltonian of the chain in this case is expressed as
\begin{equation}
{\cal H}=\sum_{n=2}^{N-1}\frac12{x'_n}^2+\sum_{n=1}^{N-1}V(x_{n+1}-x_{n}).
\label{f3}
\end{equation}
Here $\{x_n\}_{n=1}^N$ are coordinates of the rod centers.

To model the heat transfer along the chain under consideration, stochastic Langevin
thermostats are used.
The left end ($x<L_0=10$) of the chain is attached to the Langevin thermostat with temperature $T_+$,
and  the right end of the chain  with the same length [$x>(N-1)a-L_0$] --
to thermostat with temperature $T_-$.
We adopt $T_\pm=(1\pm 0.05)T$, where $T$ is average temperature of the chain.
The corresponding equations of motion have the following form:
\begin{eqnarray}
x''_n&=&-\partial {\cal H}/\partial x_n -\gamma x'_n+\xi_n^+,~~\mbox{if}~~x_n< L_0,\nonumber\\
x''_n&=&-\partial {\cal H}/\partial x_n,~~\mbox{if}~~L_0\le x_n\le (N-1)a-L_0,\label{f4}\\
x''_n&=&-\partial {\cal H}/\partial x_n -\gamma x'_n+\xi_n^-,~~\mbox{if}~~x_n> (N-1)a-L_0,\nonumber
\end{eqnarray}
where  $\gamma=0.1$ is a damping coefficient, $\xi_n^\pm$ is Gaussian white noise,
which models the interaction with the thermostats,
and is normalized as $\langle\xi_n^\pm(\tau)\rangle=0$,
$\langle\xi_n^+(\tau_1)\xi^-_k(\tau_2)\rangle=0$,
$\langle\xi_n^\pm(\tau_1)\xi_k^\pm(\tau_2)\rangle=2\gamma T_\pm\delta_{nk}\delta(\tau_2-\tau_1)$.
\begin{figure}[tb]
\includegraphics[angle=0, width=0.75\linewidth]{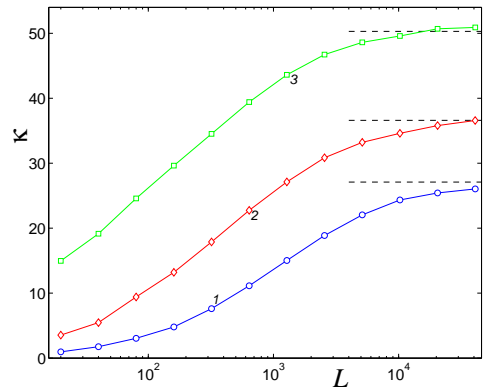}
\caption{(Color online)
Dependence of the heat conduction coefficient $\kappa$ on the distance between thermostats
$L$ for temperatures $T=0.001$ (curve 1), 0.01 (curve 2), 0.1 (curve 3) and fixed density.
The system includes rods with diameter $d=1$ with density $p=0.5$ (average distance between
the rods $a=1/p=2$), power of potential function $\alpha=5/2$.
Straight dashed lines correspond to the results obtained from Green-Kubo formula.
}
\label{fig02}
\end{figure}

System of equations (\ref{f4}) with initial conditions
${\bf X}(0)=\{x_n(0)=(n-1)a,~x'_n(0)=0\}_{n=1}^N$
was integrated numerically by Velocity Verlet method \cite{Verlet}.
Then, after some initial transient, a stationary heat flux $J$ and stationary local
temperature distribution $T(x)$ are achieved.

The total heat flux $J$ was measured as the average work produced by the
thermostats over unit time. For this sake, at each step of numerical integration $\Delta\tau$
new coordinates of the disks were calculated without account of the interaction with thermostats
${\bf X}_0(\tau+\Delta\tau)$ and then the same coordinates were calculated for chain interacting
with the thermostats, denoted as ${\bf X}(\tau+\Delta\tau)$. We define $E_+$ as the energy
of the leftmost segment of the chain which consists of disks with coordinates $x_n<L/2$
and $E_-$ as energy of the right most segment, where disks have coordinates $x_n>L/2$.
Then the work done by
the external forces in the time interval $[\tau,\tau+\Delta\tau]$ is expressed as
\begin{equation}
j_\pm=[E_\pm({\bf X}(\tau+\Delta\tau))-E_\pm({\bf X}_0(\tau+\Delta\tau))]/\Delta\tau.
\label{f5}
\end{equation}

By taking time average $J_\pm=\langle j_\pm\rangle_\tau$ we obtain the average value of energy
flux-out from the left "hot"\ thermostat and the average value of the energy flux-in into the
right "cold"\  thermostat. The value of energy flux along the chain is  $J=J_+=-J_-$.
Accuracy of this balance is considered as one of  criterions for validity of the numeric procedure.

The local heat flux, i.e. the energy flow from disk $n$ to the neighboring disk $n+1$, is defined
as $J_n=\langle j_n\rangle_\tau$, where
$$
j_n=(x_{n+1}-x_n)(x'_{n+1}+x'_n)F(x_{n+1}-x_n)/2+x'_nh_n,
$$
function $F(r)=-dV(r)/dr$, energy density distribution along the chain
$$
h_n=[{x'_n}^2+V(x_{n}-x_{n-1})+V(x_{n+1}-x_{n})]/2.
$$
(see \cite{LLP03}).

The thermal equilibrium requires all local fluxes to be equal to the total heat flux multiplied
by the chain period,  $J_n=aJ$. The fulfillment of this requirement may be considered as
a criterion for stationary regime of the heat transport.

The local temperature distribution of the chain is calculated from kinetic energy of the rod.
We divide the line segment $L$, which consists of $N$ disks, into unit-length cells $[i-1,i]$,
$i=1,...,L$ and define the following quantities:
the average number of disks in $i$-th cell is $\bar{n}_i$, and
the average kinetic energy in the cell $\bar{E}_i$.
Then the temperature of the cell is defined as $T(i)=2\bar{E}_i/\bar{n}_i$.
\begin{figure}[tb]
\includegraphics[angle=0, width=0.75\linewidth]{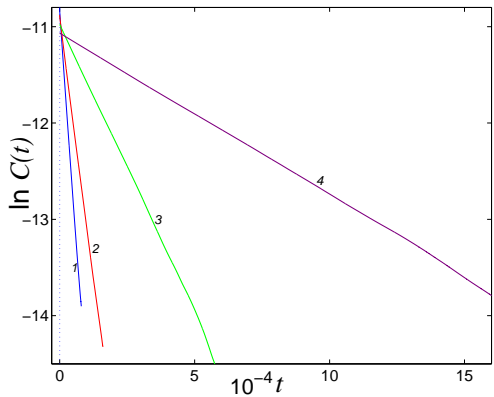}
\caption{(Color online)
Exponential decay of the autocorrelation function $C(t)$ for the chain of length $N=10000$
for temperature $T=0.033$ and density $p=0.25$, curves 1, 2, 3, 4
correspond to $\alpha=2$, 2.5, 4, 6 respectively.
}
\label{fig03}
\end{figure}

Between the thermostats we observe linear temperature gradient $T(n)$
and constant thermal flux $J$. So, the heat conduction coefficient of the free fragment
of the chain between the thermostats (of length $L-20$)
can be estimated as follows:
\begin{equation}
\kappa(L)= J[T(11)-T(L-10)]/(L-20). \label{f6}
\end{equation}

Well-known alternative way to evaluate the heat conduction coefficient is based
on well-known Green-Kubo formula
\begin{equation}
\kappa=\lim_{\tau\rightarrow\infty}\lim_{N\rightarrow\infty}\frac{1}{NT^2}\int_0^\tau C(s)ds,
\label{f7}
\end{equation}
where $C(s)=\langle J_{tot}(t)J_{tot}(t-s)\rangle_t$ is autocorrelation function of the total
heat flux in the chain with periodic boundary conditions $J_{tot}(t)=\sum_{n=1}^N j_n(t)$.

In order to compute the autocorrelation function $C(t)$ we consider a cyclic chain consisting
of $N=10^4$ particles. Initially all particles in this chain are coupled to the
Langevin thermostat with temperature $T$. After achieving
the thermal equilibrium, the system is detached from the thermostat and Hamiltonian dynamics
is simulated. To improve the accuracy, the results were averaged over $10^4$ realizations of
the initial thermal distribution.

Numeric simulation of the thermalized cyclic chain of the disks had demonstrated that
the autocorrelation function of the heat flux $C(t)$ decreases exponentially as
$t\rightarrow\infty$ -- see Fig. \ref{fig03}. Consequently, the integral in Green-Kubo formula
(\ref{f7}) converges, yielding finite value for the HCC in the chain of the disks.
Direct numeric simulation of the heat transport between the thermostats also yields saturation
of $\kappa(L)$ for large values of $L$ -- see Fig. \ref{fig02}. Both methods of simulation
yield similar values of the HCC in the saturation regime.
\begin{figure}[tb]
\includegraphics[angle=0, width=1\linewidth]{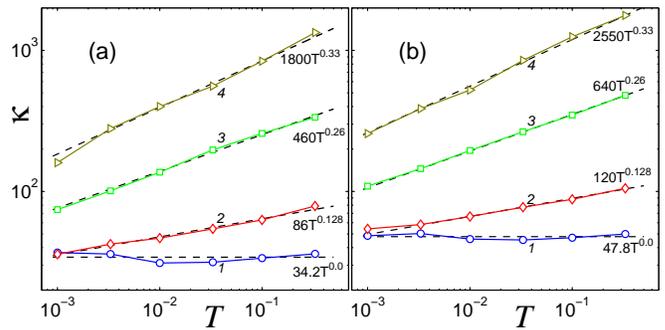}
\caption{(Color online, double logarithmic scale)
Temperature dependence of the HCC for (a) density $p=0.375$ and (b) $p=0.25$.
Curves 1, 2, 3, 4 correspond to $\alpha=2$, 2.5, 4, 6 respectively.
Straight dashed lines correspond to the power functions $\kappa=bT^\beta$.
}
\label{fig04}
\end{figure}

As it was mentioned before, we hypothesize that the observed HCC saturation may be attributed to the triple collisions
between the particles. To describe the transport process, it is convenient to define set of quasiparticles associated
with the momenta of individual particles \cite{GS04}. The quasiparticles are not affected by the double
collisions (the momenta  just hop to the next particles), but are scattered by the triple collisions.
Therefore, at phenomenological level, one can evaluate the HCC in terms of kinetic theory
in the following way:
\begin{equation}
\kappa\sim c\lambda v \sim p\lambda v.
\label{f8}
\end{equation}
Here $v$ is the characteristic velocity of the quasiparticles, $c$ is heat capacity of the system,
and $\lambda$ -- the mean free path of the quasiparticles. Scattering events are related to triple
collisions, so the mean free path corresponds to the distance traveled by the quasiparticle
between such triple collisions:
\begin{equation}
\lambda\sim 1/pP_{tr}, \label{f9}
\end{equation}
where $P_{tr}$ is the probability that given collision is triple. This probability can be estimated
as
\begin{equation}
P_{tr}\sim \tau_c/\tau_f. \label{f10}
\end{equation}
Here $\tau_f$ is the time of flight between two successive collisions, and $\tau_c$ is the
characteristic time of collision. One can also estimate $\tau_f\sim L/pv$ and therefore
\begin{equation}
\kappa\sim 1/p\tau_c. \label{f11}
\end{equation}

Evaluation of the time of collision is simple due to finite range of interaction.
If the particles collide with
relative velocity $v_0$ at infinity, then the integral of energy for the two-particle system reads as
\begin{equation}
\frac12\dot{x}^2+x^\alpha=\frac12 v_0^2, \label{f12}
\end{equation}
where $x(t)$ is the nonnegative relative displacement of the particles. Relative velocity becomes
zero at the distance $x_m=(v_0^2/2)^{1/\alpha}$. The time of collision is presented as
\begin{eqnarray}
\tau_c=2\int_{0}^{x_m}\frac{dx}{\sqrt{v_0^2-x^\alpha}}=\nonumber\\
2^{1-1/\alpha}v_0^{2/\alpha-1}\int_{0}^1\frac{d\xi}{\sqrt{1-\xi^\alpha}},~~\xi=x/x_m.
\label{f13}
\end{eqnarray}
Summarizing equations (\ref{f10}), (\ref{f11}), (\ref{f13}), and adopting $v_{0} \sim T^{1/2}$ one obtains:
\begin{equation}
\tau_c\sim v_0^{2/\alpha-1}\sim T^{1/\alpha-1/2} \Rightarrow \kappa\sim f(\alpha)p^{-1}T^{1/2-1/\alpha}
\label{f14}
\end{equation}
\begin{figure}[tb]
\includegraphics[angle=0, width=1\linewidth]{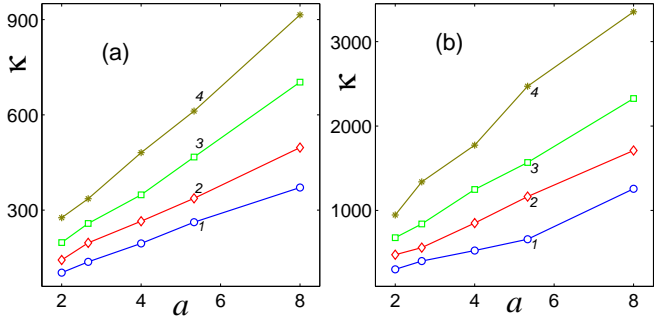}
\caption{(Color online)
Dependence of the HCC on inverse density $a=1/p$ for fixed temperature and (a)
parameter $\alpha=4$  and (b) $\alpha=6$. Curves 1, 2, 3, 4 correspond
to $T=0.01$, 0.033, 0.1, 0.33 respectively.
}
\label{fig05}
\end{figure}
\begin{figure}[tb]
\includegraphics[angle=0, width=1\linewidth]{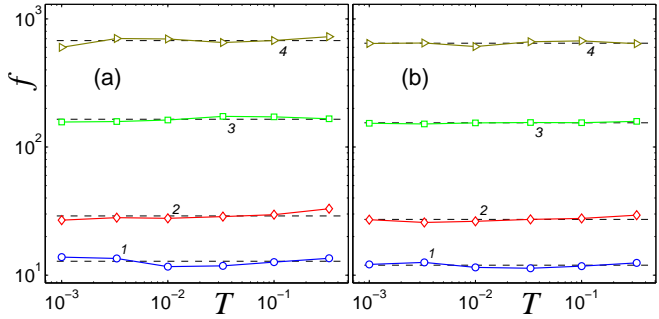}
\caption{(Color online)
Temperature dependence of the scaling function $f=p\kappa T^{1/\alpha-1/2}$ for
(a) density $p=0.375$ and (b) $p=0.25$.
Curves (1,2,3,4) correspond to $\alpha=2$, 2.5, 4, 6 respectively.
Dashed straight lines correspond to the average values of $f$.
}
\label{fig06}
\end{figure}

Equation (\ref{f14}) predicts two important features of the HCC in the considered model.
First, it has nontrivial temperature dependence. Second, quite surprisingly, it is inversely
proportional to the concentration of the particles.

Predictions for the scaling exponents in of simulations Equation (\ref{f14}) completely conform to the numeric results presented in Figures \ref{fig04} and \ref{fig05}. In the same time, scaling function $f(\alpha)$ in (\ref{f14}) remains undetermined. We can suggest that it is completely governed by intricate dynamics of the three-particle collisions and thus depends solely on exponent $\alpha$. To verify that, we plot the function $f(\alpha)=\kappa p T^{1/\alpha-1/2}$ versus exponent $\alpha$.
Numerical simulation demonstrates that the function $f(\alpha)$
does not depend on temperature $T$ and weakly depends on density $p$.
 -- see Fig.~\ref{fig06}. Figure \ref{f7} presents clear collapse of all available numeric data according to the above scaling function;
the results suggest the power law $f\approx 1.12\alpha^{3.6}$. As expected, the HCC rapidly increases as power function of $\alpha$.
\begin{figure}[tb]
\includegraphics[angle=0, width=0.95\linewidth]{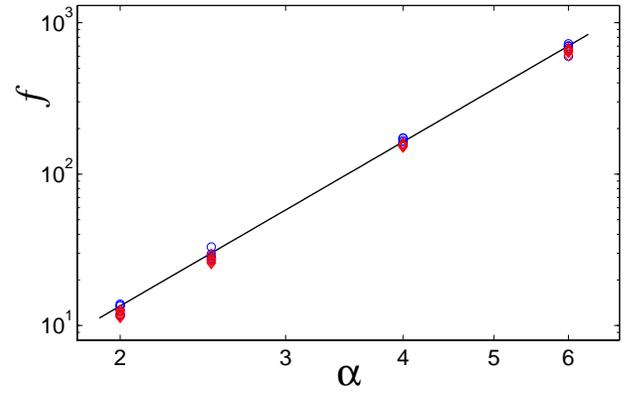}
\caption{(Color online, double logarithmic scale)
Dependence of the scaling function $f$ on power $\alpha$ for all simulated values of temperatures and particle densities.
The data collapse on straight line corresponding to $f(\alpha)=1.12\alpha^{3.6}$.
}
\label{fig07}
\end{figure}

For all explored values of $\alpha$ we observe the HCC saturation. Moreover, the observed
scaling with concentration and temperature allows concluding that the observed saturation
is caused by the triple particle collisions. Therefore, modification of the model and
removal of unphysical instantaneous collisions lead to drastic modification of the transport
properties -- namely, the observed HCC saturation. As it was mentioned above, it is not
possible to claim convergence in thermodynamic limit on the basis of numeric data
for finite system. Still, we do not observe any trend towards "resumed growth"\ of the HCC,
similar to observed in the asymmetric FPU chain \cite{FPU-anti} and billiard with
alternating masses \cite{Casati}. Of course, it might happen that simulations of even
longer chains would demonstrate such growth also in the considered chain of stiff
colliding particles. However, contrary to the asymmetric FPU and alternating-mass billiard,
the considered model also allows clear and verifiable definition of basic kinetic parameters.
The chains with possibility of dissociation possess
similar property \cite{GS14}. Intuitively, as the mean free path can be defined,
longer chains are expected to conform even better to the simple kinetic estimation
of the heat conduction coefficient. Needless to say, this latter argument also does not prove anything,
and further explorations are required to verify whether the considered model belongs to a universality class different from the FPU chain.

The authors are very grateful to Professor Stefano Lepri for useful discussions. The authors are grateful to Israel Science Foundation
(grant 838/13) for financial support.

\end{document}